\documentclass[12pt]{article}

\newcommand{\be}{\begin{equation}}
\newcommand{\ee}{\end{equation}}
\newcommand{\bea}{\begin{eqnarray}}
\newcommand{\eea}{\end{eqnarray}}
\newcommand{\beas}{\begin{eqnarray*}}
\newcommand{\eeas}{\end{eqnarray*}}

\begin{document}
\begin{titlepage}

\begin{center}

{\Large Emergence of spacetime from the algebra of total modular Hamiltonians}

\vspace{8mm}

\renewcommand\thefootnote{\mbox{$\fnsymbol{footnote}$}}
Daniel Kabat${}^{1}$\footnote{daniel.kabat@lehman.cuny.edu},
Gilad Lifschytz${}^{2}$\footnote{giladl@research.haifa.ac.il}

\vspace{4mm}

${}^1${\small \sl Department of Physics and Astronomy} \\
{\small \sl Lehman College, City University of New York, Bronx NY 10468, USA}

\vspace{2mm}

${}^2${\small \sl Department of Mathematics and} \\
{\small \sl Haifa Research Center for Theoretical Physics and Astrophysics} \\
{\small \sl University of Haifa, Haifa 31905, Israel}

\end{center}

\vspace{8mm}

\noindent
We study the action of the CFT total modular Hamiltonian on the CFT representation of bulk fields with spin. In the vacuum of the CFT the total modular Hamiltonian acts as a bulk Lie derivative, reducing on the RT surface to a boost perpendicular to the RT surface. This enables us to reconstruct bulk fields with spin from the CFT. On fields with gauge redundancies the total modular Hamiltonian acts as a bulk Lie derivative together with a compensating bulk gauge (or diffeomorphism) transformation to restore the original gauge. We consider the Lie algebra generated by the total modular Hamiltonians of all spherical CFT subregions and
define weakly-maximal Lie subalgebras as proper subalgebras containing a maximal set of total modular Hamiltonians.
In a CFT state with a bulk dual, we show that the bulk spacetime parametrizes the space of these weakly-maximal Lie subalgebras. Each such weakly-maximal Lie subalgebra induces Lorentz transformations at a particular point in the bulk manifold.  The bulk metric dual to a pure CFT state is invariant at each point under this transformation. This condition fixes the metric up to a conformal factor that can be computed from knowledge of the equation parametrizing extremal surfaces. This gives a holographic notion of the invariance of a pure CFT state under CFT modular flow. 

\end{titlepage}
\setcounter{footnote}{0}
\renewcommand\thefootnote{\mbox{\arabic{footnote}}}
\section{Introduction\label{sect:intro}}
The AdS/CFT correspondence \cite{{Maldacena:1997re}} re-packages boundary CFT properties into an effective higher-dimensional gravity theory. Understanding this equivalence from the CFT point of view has been the focus of many studies. A relationship \cite{Jafferis:2015del} which underlies the present work is the identification between the bulk total modular Hamiltonian and the CFT total modular Hamiltonian.\footnote{The total modular Hamiltonian, sometimes called the full modular Hamiltonian, acts on both
a region and its complement.} One  aspect of this identification is that the CFT total modular Hamiltonian should act on CFT representations of bulk fields in the same way that the bulk total modular Hamiltonian acts on bulk fields in an effective bulk spacetime description. From experience with weakly-coupled fields in flat space one expects that on the bulk extremal surface the action of the bulk total modular Hamiltonian should be a boost in the two dimensions perpendicular to the surface. 

For this reason the CFT representation of scalar objects localized at a bulk point should commute \cite{Kabat:2017mun, Faulkner:2017vdd} with any CFT total modular Hamiltonian whose associated bulk extremal surface (HRT surface \cite{Hubeny:2007xt}) passes through that point. This was used in \cite{Kabat:2017mun} to construct bulk operators using total modular Hamiltonians whose associated bulk extremal surfaces intersect at a bulk point. It was shown that the resulting operators agree with the complex coordinate representation of bulk operators constructed in \cite{Hamilton:2005ju, Hamilton:2006az,Hamilton:2006fh}. In the process one also gets an equation parametrizing the bulk extremal surface associated with each total CFT modular Hamiltonian. In this framework one can view the transformation property of a bulk scalar under the total modular Hamiltonian (namely that it commutes with it on the extremal surface) as a kinematic organizing principle, so perturbative corrections to the definition of a bulk scalar \cite{Kabat:2011rz} will have to obey the same condition. However $1/N$ corrections due to interactions with gauge fields and gravity \cite{Kabat:2012av, Kabat:2013wga} involve Wilson line
dressing and do not commute with the total modular Hamiltonian. One might still hope that there is an appropriate transformation law under modular flow that can be used to constrain the form of these corrections, perhaps along the lines of \cite{ Anand:2017dav, Chen:2017dnl}. For other recent uses of modular Hamiltonians in bulk reconstruction see \cite{Cotler:2017erl, Faulkner:2017tkh, Faulkner:2018faa, Czech:2018kvg}.

As explained above, the condition that a CFT operator commute with a family of total modular Hamiltonians does not give a physical bulk scalar field.\footnote{Physical in the sense of respecting an appropriate notion of bulk locality.}  But each such operator
can be associated with a point in the emergent bulk spacetime (the association is many-to-one).  We use this to show that the bulk spacetime can be identified with the space of certain
subalgebras of the Lie algebra generated by all total modular Hamiltonians.  The bulk spacetime encodes the properties and representations of these subalgebras. 

To set the stage, in section \ref{sect:Hmod} we compute the commutator of the vacuum CFT total modular Hamiltonian with CFT representations of bulk fields, both scalars and vectors. The computation shows that the commutator acts as a bulk Lie derivative. On the extremal surface associated with a given total modular Hamiltonian the commutator generates a boost in the two dimensions perpendicular to the surface. As shown in appendix \ref{appendix:massive} this condition enables one to reconstruct bulk massive vector operators from the CFT. In section \ref{sect:Hmod} we also study the action of the total modular Hamiltonian on gauge fields and metric perturbations in holographic gauge. The result is that the commutator is a Lie derivative together with a compensating gauge transformation to restore the original gauge. In section \ref{sect:emergence} we give a general discussion of how the Lie algebra generated by total modular Hamiltonians of different boundary regions is related to the emergence of the bulk manifold. In particular the bulk metric must be compatible with modular flow which constrains it up to a conformal factor.  The conformal factor can be determined from knowledge of the equation parametrizing the extremal surfaces, obtained as in \cite{Kabat:2017mun}. In section \ref{sect:example} we illustrate  these ideas for the special case of the CFT vacuum state.
Most of the computations are gathered in appendices \ref{appendix:scalar} through \ref{appendix:gravity}.

\section{Modular Hamiltonian as a bulk Lie derivative\label{sect:Hmod}}

In this section we look at properties of modular Hamiltonians for spherical regions in a CFT in its vacuum state, corresponding to an empty bulk AdS geometry.
We will find that the modular Hamiltonian acts on scalar fields and fields with spin as a bulk Lie derivative along the corresponding Killing vector, up to a compensating
gauge transformation for fields with gauge redundancy.  This can be understood as a reflection of the unbroken conformal symmetry of the vacuum state.
Many of the results in this section can only be generalized in a simple way to situations where modular flow is a local geometric operation in the bulk and boundary, for instance as in \cite{Das:2018ojl}. Nevertheless the study will be illuminating and will give some hints to the more general situation we discuss in the next section.

\subsection{Scalars}

Let us look at the modular Hamiltonian for a spherical region of radius $R$ centered around the origin in the vacuum of a CFT. It  is given by \cite{Casini:2011kv}
\be
\frac{1}{2\pi}H_{mod}=\frac{1}{2R}(Q_{0}-R^2 P_{0})
\label{mod1}
\ee
where $Q_{a}$ are the generators of special conformal transformations and $P_{a}$ are the generators of translations (see notation in Appendix \ref{appendix:scalar}).
The action of the total modular Hamiltonian on a scalar operator of dimension $\Delta$ is given by
\be
\frac{1}{2i\pi}[H_{mod}, {\cal O}(t,\vec{x})]=\frac{1}{2R}((t^2+\vec{x}^2-R^2)\partial_{t}+2t\vec{x}\partial_{\vec{x}}+2t\Delta ){\cal O}(t,\vec{x})
\ee
The bulk operator is given by \cite{Hamilton:2006az} $(\vec{x}'=\vec{X}+i\vec{y})$
\bea
&&\Phi(Z,\vec{X},T)= \frac{1}{2\Delta-d}\int dt'd\vec{y} K_{\Delta}(Z,\vec{X},T| \vec{x}',t') {\cal O}(t',\vec{x}') \nonumber\\
&&K_{\Delta} = \frac{\Gamma(\Delta-\frac{d}{2}+1)}{\pi^{d/2}\Gamma(\Delta-d+1)}\Theta\Big(\frac{Z^2+(\vec{x}' -\vec{X})^2-(t'-T)^{2}}{Z}\Big)\Big(\frac{Z^2+(\vec{x}' -\vec{X})^2-(t'-T)^{2}}{Z}\Big)^{\Delta-d}\nonumber\\
\label{bulkscalar}
\eea
where $\Theta(x)$ is the step function.
A computation in appendix \ref{appendix:scalar} gives
\begin{equation}
\frac{1}{2i\pi}[H_{mod},\Phi(Z,\vec{X},T)]=\frac{1}{2R}(Z^2+\vec{X}^2-R^2+T^2)\partial_{T}\Phi(Z,\vec{X},T)+\frac{1}{R}(TZ\partial_{Z}+T\vec{X}\partial_{\vec{X}})\Phi(Z,\vec{X},T)
\end{equation}
If we label the vector field $(\xi^z,\xi^{\vec{X}},\xi^T)$ and define
\begin{equation}
\xi_{R,0}^{\mu}=( \frac{1}{R}TZ, \frac{1}{R}T\vec{X}, \frac{1}{2R}(Z^2+\vec{X}^2-R^2+T^2))
\end{equation}
then
\begin{equation}
\frac{1}{2i\pi}[H_{mod},\Phi(Z,\vec{X},T)]=\xi_{R,0}^{\mu}\partial_{\mu} \Phi(Z,\vec{X},T).
\label{xiphi}
\end{equation}
On the RT surface $(Z^2+\vec{X}^{2}=R^2, \ \ T=0)$, the vector field vanishes and one has (see also \cite{Faulkner:2017vdd})
\be
[H_{mod},\Phi]=0
\label{basic1}
\ee
More generally for the total modular Hamiltonian of a spherical region of radius $R$ centered around $Y_i$, $i=1,\cdots d$, equation (\ref{xiphi}) still holds with
\begin{equation}
\xi_{R,Y_{i}}^{\mu}=\Big(\frac{1}{R}TZ, \frac{1}{R}T(\vec{X}-\vec{Y}), \frac{1}{2R}(Z^2+(\vec{X}-\vec{Y})^2-R^2+T^2)\Big)
\label{genxi}
\end{equation}
In \cite{Kabat:2017mun}  it was shown that one can use (\ref{basic1}) to construct a bulk scalar operator in AdS${}_3$, by demanding that the operator $\Phi$ obeys (\ref{basic1}) for two different modular Hamiltonians based on two different segments of the boundary. Similar calculations can be done in AdS${}_{d+1}$,
by demanding that $\Phi$ commutes with $d$ different total modular Hamiltonians based on different spherical regions of the boundary. As a byproduct of the solution one also gets a parametrization of the RT surface \cite{Ryu:2006bv} in these coordinates \cite{Kabat:2017mun}, namely
\be
Z^2+(\vec{X}-\vec{Y})^2=R^2 \quad {\rm and} \quad T = 0.
\ee
Note that when solving (\ref{basic1}) one only finds solutions up to a scalar function of the spacetime coordinates.\footnote{If we could demand that a result of the form
(\ref{xiphi}) be satisfied throughout the spacetime for every total modular Hamiltonian, then we could have fixed the overall spacetime coefficient and by that also fix the
$\xi^{\mu}$'s. This works for the vacuum state of the CFT but not in general.}

\subsection{Massive vectors\label{sect:MassiveVectorHmod}}

We can now compute the action of the CFT total modular Hamiltonian on the CFT representation of bulk vectors.

The total modular Hamiltonian for a spherical region centered around the origin is given by (\ref{mod1}).
The CFT representation of bulk massive vectors starts with a  non-conserved primary current $j_{\mu}$ of dimension $\Delta$. We label $j_{z}=\frac{1}{d-1-\Delta}(-\partial_{0}j_{0}+\partial_{i}j_{i})$. The bulk massive vector fields are then given by \cite{Kabat:2012hp}
\begin{eqnarray}
ZV_{\mu} &= & \int K_{\Delta}\ j_{\mu} +\frac{Z}{2(\Delta-\frac{d}{2}+1)}\partial_{\mu}\int K_{\Delta+1}\  j_{z}\nonumber\\
V_{z} &=& \int K_{\Delta} \  j_{z}
\label{massivevectors}
\end{eqnarray}
where $K_{\Delta}$ is given in (\ref{bulkscalar}).
Computing the action of the total modular Hamiltonian on this expression we get (see appendix \ref{appendix:massiveHmod})
\bea
\frac{1}{2i\pi}[ H_{mod}, V_{0}(Z,\vec{x},T) ] &=& \xi_{R,0}^{\mu}\partial_{\mu}V_{0}+\frac{Z}{R}V_{z}+\frac{\vec{X}\cdot \vec{V}}{R}+\frac{T}{R}V_{0} \nonumber\\
\frac{1}{2i\pi}[ H_{mod},V_{i} ] &=& \xi_{R,0}^{\mu}\partial_{\mu}V_{i} +\frac{X_{i}}{R}V_{0}+\frac{T}{R}V_{i} \label{hmodvimass} \\
\frac{1}{2i\pi}[ H_{mod},V_{Z} ] &=&\xi_{R,0}^{\mu}\partial_{\mu}V_{Z}+  \frac{Z}{R}V_{0}+\frac{T}{R}V_{z} \nonumber
\eea
This can be written as 
\begin{equation}
\frac{1}{2i\pi}[ H_{mod}, V_{\nu}(Z,\vec{X},T) ]=\xi_{R,0}^{\mu}\partial_{\mu}V_{\nu}+V_{\mu} \partial_{\nu}\xi_{R,0}^{\mu}\equiv ({\cal L}_{\xi} V)_{\nu}.
\label{xivi}
\end{equation}
Thus the action of the total modular Hamiltonian is just a bulk Lie derivative.
On the RT surface where the vector field $\xi^{\mu}_{R,0}$ vanishes and $T=0$, we can write this as 
\be
\frac{1}{2i\pi}[ H_{mod}, V_{\perp} \pm V_{0} ] =\pm(V_{\perp}\pm V_{0}), \qquad
[H_{mod},V_{||}] = 0
\label{modvigen}
\ee
where $V_{\perp}, V_{||}$ are the components perpendicular and parallel to the corresponding RT surface. Thus we see that, as expected, the CFT total modular Hamiltonian acts on bulk fields on the RT surface (represented as CFT operators) as a boost in the two dimensions perpendicular to the RT surface. Equation (\ref{modvigen}) can be used to obtain the CFT representation of a bulk massive vector field in a manner similar to the scalar case \cite{Kabat:2017mun}. This is done in Appendix \ref{appendix:massive}.

\subsection{Gauge fields}

As we saw, the action of the vacuum CFT total modular Hamiltonian on vector fields is given by (\ref{hmodvimass}).
Acting on gauge fields in the bulk in the gauge $A_{Z}=0$, one has to combine the boost with a compensating gauge transformation to restore $A_{Z}=0$ gauge. The combined action should then be
\bea
\frac{1}{2i\pi}[ H_{mod}, A_{0}(Z,\vec{X},T) ] &=& \xi_{R,0}^{\mu}\partial_{\mu}A_{0}+\frac{T}{R}A_{0}+\frac{\vec{X}\cdot \vec{A}}{R}-\partial_{0} \lambda \\
\frac{1}{2i\pi}[ H_{mod},A_{i}(Z,\vec{X},T ] &=&\xi_{R,0}^{\mu}\partial_{\mu}A_{i} +\frac{T}{R}A_{i}+ \frac{X_{i}}{R}A_{0} -\partial_{i} \lambda\nonumber
\eea
where
\be
\partial_{z} \lambda = \frac{Z}{R}A_{0}\,,
\ee
or in condensed notation ($a=0,\cdots, d-1$)
\be
\frac{1}{2i\pi}[ H_{mod}, A_{a}(Z,\vec{X},T) ] = ({\cal L}_{\xi} A)_{a}|_{A_{z}=0}-\partial_{a} \lambda\,.
\ee
One might get worried: the CFT total modular Hamiltonian does not know which gauge we are in or even that there is a bulk gauge freedom, so how could it possibly reproduce this? The point is that in the CFT it is the CFT representation of a bulk gauge field which depends on the gauge choice while the CFT total modular Hamiltonian is fixed.

For example the representation of a bulk gauge field in AdS${}_{d+1}$ in $A_Z=0$ gauge is \cite{Kabat:2012hp}, with $\vec{x}'=\vec{X}+i\vec{y}$ 
\bea
ZA_{0}(Z,\vec{X},T)=\frac{1}{{\rm vol}(S^{d-1})}\int dt'd\vec{y}\,\delta \left ( \frac{Z^2+(\vec{x}' -\vec{X})^2-(t'-T)^{2}}{2Z} \right )j_{0}(t',\vec{x}') \nonumber\\
ZA_{i}(Z,\vec{X},T)=\frac{1}{{\rm vol}(S^{d-1})}\int dt'd\vec{y}\,\delta \left ( \frac{Z^2+(\vec{x}' -\vec{X})^2-(t'-T)^{2}}{2Z} \right )j_{i}(t',\vec{x}')
\eea
We compute  in  appendix \ref{appendix:gauge} $[ H_{mod}, A_{a}(Z,\vec{X},T) ] $ and find as expected
\be
\frac{1}{2i\pi}[ H_{mod}, A_{a}(Z,\vec{X},T) ] = ({\cal L}_{\xi} A)_{a}|_{A_{z}=0}-\partial_{a} \lambda
\ee
So also in this case the CFT total modular Hamiltonian reproduces the correct result. Note that on the RT surface this is just a boost perpendicular to the RT surface followed by a compensating gauge transformation to restore the original $A_{Z}=0$ gauge.

\subsection{Gravity}

The expression for a bulk metric perturbation in holographic gauge $h_{ZZ}=h_{Za}=0$ (where $a,b$ range over $0,1,\cdots,d-1$) is \cite{Kabat:2012hp}
\begin{equation}
Z^2h_{ab}=\frac{d\Gamma(d/2)}{2\pi^{d/2}}\int dt'd\vec{y}'\,\Theta \left ( \frac{Z^2+(\vec{x}' -\vec{X})^2-(t'-T)^{2}}{2Z} \right )T_{ab}(t',\vec{x}')
\label{bulkgrav}
\end{equation}
One can compute the action of the CFT total modular Hamiltonian on bulk gravitons to be (see appendix \ref{appendix:gravity})
\bea
\frac{1}{2i\pi}[ H_{mod}, h_{ij}(Z,\vec{X}) ] &=&\xi^{\mu}_{R,0}\partial_{\mu}h_{ij}+\frac{2T}{R}h_{ij}+ \frac{X_{i}}{R}h_{0j}+\frac{X_{j}}{R}h_{i0}-\frac{1}{2R Z^2}(\partial_{i} \epsilon_j +\partial_{j} \epsilon_{i}) \nonumber\\
\frac{1}{2i\pi}[ H_{mod},h_{0i} (Z,\vec{X})] &=&  \xi^{\mu}_{R,0}\partial_{\mu}h_{0i}+\frac{2T}{R}h_{0i}+\frac{X^{j}}{R}h_{ji}+\frac{X_{i}}{R}h_{00}-\frac{1}{2R Z^2}(\partial_{i} \epsilon_0 +\partial_{0} \epsilon_{i})\nonumber \\
\frac{1}{2i\pi}[ H_{mod},h_{00} (Z,\vec{X})] &=& \xi^{\mu}_{R,0}\partial_{\mu}h_{00}+\frac{2T}{R}h_{00}+ 2\frac{X^{j}}{R}h_{j0}+\frac{X_{i}}{R}h_{00}-\frac{2}{2R Z^2}\partial_{0} \epsilon_0
\label{achmodgravbulk}
\eea
where 
\begin{equation}
\epsilon_{a}=\frac{d\Gamma(d/2)}{2\pi^{d/2}}\int dt'd\vec{y}'\Theta \left ( \frac{Z^2+(\vec{x}' -\vec{X})^2-(t'-T)^{2}}{2Z} \right )(Z^2+(\vec{x}' -\vec{x})^2-(t'-T)^{2})T_{0 a}(t',\vec{x}')
\end{equation}
parametrizes a diffeomorphism which satisfies
\begin{equation}
\frac{1}{Z^2}\partial_{Z} \epsilon_{a}=2Z h_{0a}
\end{equation}
and thus restores holographic gauge after the boost.
In condensed notation $(\xi^{\mu}=\xi^{\mu}_{R,0})$
\be
\frac{1}{2i\pi}[ H_{mod}, h_{ab}(Z,\vec{X}) ]  =(\xi^{\mu}\partial_{\mu} h_{ab}+ \partial_{a}\xi^{\mu}h_{\mu b}+ \partial_{b}\xi^{\mu}h_{a \mu})|_{h_{ZZ}=h_{Zc}=0}-\frac{1}{2RZ^2}(\partial_{a}\epsilon_{b}+\partial_{b}\epsilon_{a})
\ee
The first term on the right is a Lie derivative evaluated in holographic gauge, $({\cal L}_{\xi} h_{ab})|_{h_{ZZ}=h_{Zc}=0}$, while the second term is a diffeomorphism restoring holographic gauge.
On the RT surface ($\xi_{R,0}^{\mu}=0$, $T=0$) we again get the expected result of a boost perpendicular to the RT surface plus a compensating
diffeomorphism.

\section{Emergence of the bulk spacetime\label{sect:emergence}}

In this section we describe how the bulk spacetime (dual to some state $|\Psi\rangle$ in the CFT) arises from considerations involving CFT total modular Hamiltonians. We do this by considering the Lie algebra generated by the total modular Hamiltonians of the CFT state for different regions and their representations.

Take the set of total modular Hamiltonians (appropriate for the state $|\Psi\rangle$) associated with spherical regions in the CFT (this is not crucial, one can pick any other fixed shape). We label these total modular Hamiltonians by $d+1$ parameters, say the centers of the spheres $Y_{i}$, $i=0,\ldots ,d-1$ (including time) and their spatial radii $R$.  We start with this set  and generate by repeated use of commutators\footnote{Generally the commutator of two total modular Hamiltonians is not the total modular Hamiltonian of any region.} and linear combinations a Lie algebra $A^{\Psi}$ (probably infinite-dimensional in the general case). All members of the algebra annihilate the state, $A^{\Psi} |\Psi\rangle=0$. We then look for what we call weakly-maximal Lie subalgebras $G^{\Psi}_{P}$ of $A^{\Psi}$. Weakly-maximal means that $G^{\Psi}_{P}$ is a proper subalgebra that contains the largest possible number of total modular Hamiltonians associated with spherical CFT subregions.  The label $P$ parametrizes
these Lie subalgebras if they exist. From now on we drop the label $\Psi$ for convenience.

One way to define subalgebras is to look for the objects (CFT operators) which they leave invariant.
Thus given a modular Hamiltonian $H_{mod}(Y_i,R)$ we look for operators in the CFT  which are solutions to the equation\footnote{One could relax this condition and
only require that the commutator vanish inside a code subspace.  In the analysis that follows it does not seem that anything is gained by this generalization.}
\begin{equation}
[H_{mod}(Y_i,R),\Phi]=0\,.
\label{modsol}
\end{equation}
Note that in a holographic theory bulk operators that live on the extremal bulk surface will obey this condition \cite{Kabat:2017mun, Faulkner:2017vdd}.  It is possible for $\Phi$ to commute with more than one modular Hamiltonian.  From the bulk perspective this happens if the extremal
surfaces intersect.  Given such a $\Phi$ we define $H_{\Phi}$ as the set of total modular Hamiltonians that leave $\Phi$ invariant.\footnote{As shown in \cite{Kabat:2018pbj}
$\Phi$ may not be a completely kosher operator.  But $|\Phi\rangle \equiv \Phi|\Psi\rangle$ is a well-defined state, so strictly speaking we should define $H_\Phi$ as the set of modular Hamiltonians that annihilate $\vert\Phi\rangle$.  For notational convenience we will overlook this subtlety.}

Now let's count parameters.  It's simplest to work on a fixed-time slice of AdS${}_{d+1}$ / CFT${}_d$.  Modular Hamiltonians for spherical regions are then labeled by $d$ parameters, the centers of the spheres and their radii, and RT surfaces have codimension 1.  Requiring that a given bulk point lies on an extremal surface is therefore one condition on $d$ parameters, so we expect
a $(d-1)$-parameter family of modular Hamiltonians that leave the point invariant.  We are interested in weakly-maximal subalgebras so
this is the case we will consider: we expect an operator $\Phi$ associated with a bulk point to be invariant under a $(d-1)$-parameter family of total modular Hamiltonians.\footnote{CFT operators associated not with points but with higher-dimension regions of the bulk could be invariant under smaller families of modular Hamiltonians.}  Moreover since RT surfaces are codimension 1 it
generically takes $d$ modular Hamiltonians to specify a bulk point.  This seems like $d^2$ parameters, but $d(d-1)$ of these parameters are redundant and correspond to the same bulk point.
So not surprisingly we find a $d$-parameter family of bulk points on a spatial slice.

The operator $\Phi$ is associated with a bulk point.  But many other operators are associated with the same point, since for example we can build a bulk scalar from any spinless primary operator
in the CFT.  All these operators will be invariant under the same $H_\Phi$, so what is uniquely associated with a bulk point is $H_{\Phi}$.  Thus we can regard the bulk spacetime as the space of $H_{\Phi}$.
Restoring time, we expect the space of $H_{\Phi}$ to have $d+1$ parameters.  Let's call these parameters $(\vec{X},Z,T)$.  These parameters define a coordinate system for the bulk spacetime.
We will use these parameters both to label the sets $H_{\vec{X},Z,T}$ and to label operators $\Phi(\vec{X},Z,T)$ that are invariant under $H_{\vec{X},Z,T}$.

At this stage each point of the bulk manifold is associated with a set $H_{\vec{X},Z,T}$.
The condition (\ref{modsol}) guaranties that taking commutators and linear combinations will turn this set into a Lie subalgebra $G_{\vec{X},Z,T}$ that leaves $\Phi(\vec{X},Z,T)$ invariant,
\begin{equation}
[G_{\vec{X},Z,T},\Phi(\vec{X},Z,T)]=0.
\end{equation}
We conjecture that these subalgebras are weakly maximal and that all weakly-maximal subalgebras are produced in this way.\footnote{Our results do not rely on this conjecture, since in practice it is the property of leaving $\Phi$ invariant, not the maximality, which will be important in what follows.}
The smoothness of the bulk manifold is inherited from the smoothness of the space of total modular Hamiltonians with respect to its parameters $(Y_i,R)$.  Bulk coordinate transformations are just a re-labeling of the weakly-maximal subalgebras.

A note of caution. All these considerations are appropriate for theories with a holographic dual. Theories where the above structure and properties of the total modular Hamiltonians are not present do not have a holographic dual. Even theories which do have the above properties are not guaranteed to have a useful dual since it is not guaranteed that there is a macroscopic bulk with a well-defined low-energy theory.  We would also like to stress that operators $\Phi(\vec{X},Z,T)$ which satisfy (\ref{modsol}) are not CFT representations of physical bulk scalar fields: scalar fields interacting with gravity (or gauge fields) do not commute with the modular Hamiltonian, rather they only commute with it to leading order in $1/N$.\footnote{It is however possible that the action of the modular Hamiltonian on physical bulk fields can be kinematically constrained (as in \cite{ Anand:2017dav}) and used as a guiding principle for arranging a perturbative expansion for bulk fields.} So we are not trying to construct physical bulk fields. Instead we are only using solutions of (\ref{modsol}) to parametrize the bulk spacetime.

In a holographic theory the bulk metric represents in some way the CFT state $|\Psi\rangle$. The pure state $|\Psi\rangle$ is annihilated by the total modular Hamiltonian of any subregion (that is, $|\Psi\rangle$ is invariant under modular flow). Some expression of this invariance should apply to the metric. But in general modular flow is not geometric so what should we expect? To see what happens
we look at the representations of the weakly-maximal Lie subalgebras  $G_{\vec{X},Z,T}$ on the CFT Hilbert space. On the extremal surface the action of the total modular Hamiltonian on CFT representations of bulk fields is expected to be a boost in the two dimensions normal to the surface. We saw an
example of this in section \ref{sect:MassiveVectorHmod}, where for a bulk vector field and the CFT vacuum we found operators ($V_{\pm}, V_{||}$) in the CFT such that on the extremal surface
\be
\frac{1}{2i\pi}[ H_{mod}, V_{\pm} ] =\pm V_{\pm},  \ \ 
[H_{mod},V_{||}] = 0
\ee
More generally we expect
\begin{equation}
[H_{mod},V_{\mu}(\vec{X},Z,T)]|_{\hbox{\small extremal surface}} = \Lambda_\mu{}^{\nu}V_{\nu}(\vec{X},Z,T)|_{\hbox{\small extremal surface}}
\label{modvecb}
\end{equation}
where $\Lambda_\mu{}^{\nu}(\vec{X},Z,T)$ is a matrix representation of a Lorentz boost generator.  One could work out the explicit representation of these boost generators by constructing a bulk scalar $\Phi(\vec{X},Z,T)$ and evaluating
$[H_{mod},\partial_\mu \Phi]$, as explained in the paragraph below (\ref{modactspe}).  In carrying out this construction note that, having chosen parameters $\vec{X},Z,T$ on the space of weakly-maximal subalgebras, we are now using
these parameters to define a coordinate basis for the tangent space.  

While modular Hamiltonians are mapped to boosts, when acting on CFT representations of bulk fields with spin their commutators are mapped to rotations.
Thus there is a map (usually many to one) from $G_{\vec{X},Z,T}$  to the Lorentz algebra. This structure gives rise to a ``local'' Lorentz algebra at each bulk spacetime point. This is of course not the usual local symmetry of the tetrad formalism: each element of $G_{\vec{X},Z,T}$ induces a rigid transformation (generically non-geometric) throughout the spacetime and is not a gauge symmetry.  But it still reduces to the Lorentz algebra at the associated bulk point.\footnote{One may think of the modular Hamiltonians as acting on Lorentz indices of bulk fields in a  fixed choice for the non-coordinate (tetrad) basis. This also provides a way to define fermions on curved space, and one would expect the CFT total modular Hamiltonian to act on the Lorentz indices of the CFT representation of bulk fermions.}

Hence the action of the  modular Hamiltonian on  fields with spin on the extremal surface is geometric as in (\ref{modvecb}). One can expect that the metric at each point will be  invariant in the sense that
\begin{equation}
\Lambda_{\mu}{}^{\alpha}(\vec{X},Z,T)g_{\alpha \nu}(\vec{X},Z,T)+ \Lambda_{\nu}{}^{\alpha}(\vec{X},Z,T)g_{\mu \alpha}(\vec{X},Z,T)=0
\label{liederg}
\end{equation}
where $ \Lambda_{\mu}{}^{\alpha}(\vec{X},Z,T)$ are Lorentz generators corresponding to an element of $G_{\vec{X},Z,T}$. This equation fixes the metric at each point up to a conformal factor.
This can be understood as follows. Given a codimension-2 spacelike surface we can parametrize spacetime by Gaussian normal coordinates $(q^i, x^{\pm})$. The surface is at $x^{\pm}=0$ and the $q^i$ parametrize the surface. The metric near the surface has the form
\be
ds^2=dx^{+}dx^{-}+\gamma_{ij}dq^{i}dq^{j}+O(x^\pm).
\ee
Thus the metric on the surface is invariant under a boost in the $x^{\pm}$ directions. Given a point it has many surfaces going through it and the matrix $\Lambda_\mu{}^{\nu}$ in (\ref{liederg}) encodes the relationship between the different boosts at the same spacetime point, i.e.\ the relationship between the normal and parallel directions of the different surfaces. Thus one has the angles between the different normal and parallel vectors which fixes the metric up to a conformal factor.

Another avenue to define a metric is to look for a natural metric on the set of Lie subalgebras $\{G_{\vec{X},Z,T}\}$. Since to each member in this set there is an associated CFT state $|\Phi(\vec{X},Z,T)\rangle$, we can use the overlap of theses states as a measure of the distance between the Lie subalgebras. If the overlap was not divergent we could have tried to use the Fubini-Study metric as in \cite{Miyaji:2015fia}, and this is still possible if one finds a natural regularization.   Instead as in \cite{Roy:2018ehv} we use the singular limit as a measure for the metric, which is aided by a natural identification of the overlap (in the leading $1/N$ expansion) with the CFT two-point function of local bulk scalars. Then as $(\vec{X},Z,T) \rightarrow (\vec{X}',Z',T')$ the overlap will behave as 
\begin{equation}
\langle\Phi(\vec{X},Z,T)|\Phi(\vec{X}',Z',T')\rangle \rightarrow \frac{a}{\sigma^{d-1}} 
\label{bulksing}
\end{equation}
where $\sigma^2=g_{\mu \nu}(x-x')^{\mu}(x-x')^{\nu}$ and $a$ is some constant. However since solutions to (\ref{modsol}) are determined only up to an overall position-dependent coefficient, the metric cannot be uniquely extracted from the singularity structure. Only the metric up to a conformal factor can be extracted from the expected singularity.\footnote{The conformal factor can be deduced by introducing additional assumptions.  See \cite{Roy:2018ehv}.} As we show now these two notions of the metric, under some reasonable assumptions,  are compatible.
 
The modular Hamiltonian acting on a bulk scalar field is expected to induce a non-local transformation on the scalar field. However it seems natural (and can be seen in some simple cases) that near the extremal surface it still obeys\footnote{Similar considerations were used in \cite{Jacobson:2015hqa, Faulkner:2017vdd}, see also \cite{Arias:2018tmw}.}
\begin{equation}
[H_{mod}, \Phi(\vec{X},Z,T)]=\xi^{\mu}\partial_{\mu} \Phi (\vec{X},Z,T)+ {\rm \ less\ singular}
\label{modactspe}
\end{equation}
where ``less singular'' means that the Lie derivative part  is the leading contribution near the extremal surface when evaluated inside a two-point function with another scalar on the extremal surface.
In addition we expect that the Lorentz boost matrix associated with the total modular Hamiltonian is given by $\Lambda_\mu{}^{\nu} =\partial_{\mu} \xi^{\nu}|_{\rm extremal\ surface}$. Using (\ref{modactspe}) in a two-point function with a scalar field $\Phi(X'_{i},Z',T')$ that sits exactly on the extremal surface one gets
\begin{eqnarray}
0 &=& \langle\Psi| \Phi(X_{i},Z,T)[H_{mod},\Phi(X'_{i},Z',T')]|\Psi\rangle=\langle\Psi|[\Phi(X_{i},Z,T),H_{mod}]\Phi(X'_{i},Z',T')|\Psi\rangle\nonumber\\
&=& -\xi^{\nu}\partial_{\nu}  \langle\Psi|\Phi(X_{i},Z,T)\Phi(X'_{i},Z',T')|\Psi\rangle+ {\rm \ less\ singular}
\label{bulksing2}
\end{eqnarray}
Labeling the difference of bulk coordinates as $(x-x')^{\mu}$ and using
\be
\xi^{\nu}=\Lambda^{\nu}_{\mu}(x-x')^{\mu}+O((x-x')^{2})
\ee
and also (\ref{bulksing}) we get
\be
\Lambda_{\mu}{}^{\nu}g_{\nu \beta}(x-x')^{\beta}(x-x')^{\mu}=0
\ee
where  $g_{\alpha \beta}$ and $\Lambda_{\mu}{}^{\nu}$ are evaluated at the point on the extremal surface $(X'_{i},Z',T')$. Since this is true for any $(x-x')^{\alpha}$ one gets on the extremal surface,
\begin{equation}
\Lambda_{\mu}{}^{\alpha}g_{\alpha \nu}+ \Lambda_{\nu}{}^{\alpha}g_{\mu \alpha}=({\cal L}_{\xi} g_{\mu\nu})|_{\rm extremal\  surface}=0
\label{MetricInvariant}
\end{equation}
which is just (\ref{liederg}).

We saw that the metric, up to a conformal factor, can be extracted from the structure and representations of the weakly-maximal subalgebras. To get the conformal factor
we can proceed as follows.  There are special bulk surfaces associated with each total modular Hamiltonian. These are codimension-2 surfaces with the property that they are made from a $(d-1)$-dimensional continuous family of points $(\vec{X},Z,T)$, such that the given total modular Hamiltonian $H_{mod}$ is a member of all the weakly-maximal subalgebras $G_{\vec{X},Z,T}$ appearing in that family.  These surfaces are identified with the bulk extremal surfaces that intersect the AdS boundary on the boundary of the region associated with the given $H_{mod}$.
 
To fix the conformal factor we require that the extremal surfaces deduced using the modular Hamiltonians have vanishing mean curvature (that is, the trace of their extrinsic curvature is zero). Since each surface has codimension 2 it has two independent normal vectors and thus has two mean curvatures $h_{(1)}$, $h_{(2)}$.
We can choose the normal vectors $n^{\alpha}_{(i)}$ to be orthogonal to each other.
Under a conformal transformation $\tilde{g}_{\mu \nu }= e^{2f}g_{\mu \nu}$ the mean curvature of the surface transforms to
\begin{equation}
\tilde{h}_{(i)}=e^{-f}\left(h_{(i)}+(d-1)n^{\alpha}_{(i)}\partial_{\alpha} f \right)
\label{extremcon}
\end{equation}
where $n_{\alpha}^{(i)}$ are the unit normal vectors  to the surface in the original metric and $(d-1)$ is the dimension of the surface. 
This determines the normal derivatives of the conformal factor for each extremal surface passing through the bulk point, which should be enough to fix the conformal factor uniquely.\footnote{There may not be a solution to these equations in which case there is no bulk spacetime.}

\subsection{Special case: CFT vacuum\label{sect:example}}

In this section we show how the reconstruction procedure of section 3 works for the vacuum state of the CFT.  Beginning from the CFT vacuum we derive the algebra of modular Hamiltonians of different spherical regions and obtain the corresponding bulk metric.  For simplicity we work in AdS${}_3$ but identical conclusions hold in higher dimensions.

For the vacuum state of a two-dimensional CFT the total modular Hamiltonian for a segment $(y_1,y_2)$ of the boundary at time $T=0$ is (see appendix A for conventions for conformal generators)
\begin{equation}
\frac{1}{2\pi}H_{mod}^{1,2}=\frac{1}{y_2-y_1}(Q_{0}+y_1 y_2 P_{0}+(y_1 + y_2)M_{01}).
\label{HmodAdS3}
\end{equation}
We can identify modular Hamiltonians whose extremal surfaces intersect by looking for solutions to the equations
\begin{equation}
[H_{mod}^{1,2},\Phi] = [H_{mod}^{3,4},\Phi] = 0
\label{hmodphi}
\end{equation}
This was done in \cite{Kabat:2017mun} by explicitly constructing the operator $\Phi$, with the result that the two extremal surfaces intersect at a bulk point $(Z_0,X_0,T=0)$ implicitly determined by the conditions\footnote{If the RT surfaces do not intersect then $Z_0$ becomes imaginary.}
\begin{equation}
Z_0^2 = (y_2-X_{0})(X_{0}-y_1)=(y_4-X_{0})(X_{0}-y_3)
\label{Intersect}
\end{equation}
We can think about this result in two ways.  If we imagine holding $Z_0$ and $X_0$ fixed, we see that (\ref{HmodAdS3}) defines a one-parameter family of intersecting
modular Hamiltonians provided the parameters $y_1$ and $y_2$ are related by $Z_0^2 = (y_2 - X_0)(X_0 - y_1)$.  On the other hand we can hold $y_1$ and $y_2$ fixed.
Then we can read off the RT surface associated with $H_{mod}^{1,2}$, namely the bulk semicircle $Z^2 = (y_2 - X)(X - y_1)$.

The commutator of two total modular Hamiltonians is
\begin{equation}
\frac{1}{4\pi^2}[H_{mod}^{1,2},H_{mod}^{3,4}]= -i\alpha \left( Q_1 +2X_{0} D+(Z_0^2+X_{0}^{2})P_1\right)\equiv2 i \alpha Z_0 J
\label{defj}
\end{equation}
where $X_0$ and $Z_0$ are determined by (\ref{Intersect}) and
\begin{equation}
\alpha = \frac{y_3+y_4-(y_1 +y_2)}{(y_2-y_1)(y_4-y_3)}
\end{equation}
Note that in AdS${}_3$ any two modular Hamiltonians whose RT surfaces intersect at the same point have the same commutator up to an overall coefficient.  Given that there is only one rotation generator in $2+1$ dimensions, this is consistent with the expectation that modular Hamiltonians generate boosts about the RT surface with a commutator that is proportional to a rotation $J$ about the intersection point.\footnote{This expectation can be explicitly verified by computing $[J,V_{\mu}]$ which shows that $J$  acts as a rotation in the $(Z,X)$ plane.} 

To put the algebra in a standard form we allow $y_1$ and $y_2$ to be arbitrary and define $K_1 = \frac{1}{2\pi}H_{mod}^{1,2}$.  With $(Z_0,X_{0})$ corresponding to some point on the extremal surface associated with $H_{mod}^{1,2}$ we define
$K_2=\frac{1}{2\pi}H_{mod}^{3,4}$ with
\begin{eqnarray}
y_3=\frac{1}{2X_0 -(y_1+y_2)}(2(Z_0^2+X_{0}^{2})-X_{0}(y_1+y_2)-Z_0(y_2-y_1)) \nonumber \\
y_4=\frac{1}{2X_0 -(y_1+y_2)}(2(Z_0^2+X_{0}^{2})-X_{0}(y_1+y_2)+Z_0(y_2-y_1)) \nonumber
\end{eqnarray}
Using $J$ from (\ref{defj}) with parameters $(Z_0,X_{0})$ one finds
\begin{equation}
[K_1,J] = iK_2, \ \ [K_{2},J]= -i K_1, \ \ [K_1,K_2]= iJ
\end{equation}
Thus starting from modular Hamiltonians whose RT surfaces intersect at a point, we obtain through commutators and linear combinations all generators of the Lie algebra $so(2,1)$. This Lie algebra can be exponentiated to $SO(2, 1)$, which becomes in this case the stabilizer group that leaves the intersection point invariant. As shown in section 2, at each point of the RT surface the CFT operators that correspond to bulk scalar and bulk vector fields form a representation of the associated Lorentz algebra.

If we started with all possible total modular Hamiltonians based on single segments (i.e.\ whose RT surfaces do not necessarily intersect at a common point) then it is easy to see that we would get a Lie algebra spanned by six independent generators $P_0$, $P_1$, $Q_0$, $Q_1$, $D$, $M_{01}$. These generate the Lie algebra $so(2, 2)$, and by exponentiating we get the group $SO(2, 2)$ which in this case can be identified with the isometry group of AdS${}_3$.  Thus $so(2,1)$ is a weakly-maximal Lie subalgebra of $so(2,2)$.  As discussed in general in section 3, each bulk point is associated with such a weakly-maximal Lie subalgebra.

In the special case of the vacuum state of the CFT, the weakly-maximal  Lie subalgebras associated with different bulk points are isomorphic: they are related by conjugating
by an element of $SO(2,2)$.  This can be traced to the fact that in empty AdS, given any $\Phi(Z,X,T)$ and $\Phi(Z',X',T')$ which are invariant in the sense of (\ref{hmodphi}),
there is an element $g$ of $SO(2,2)$ such that $g\Phi(Z,X,T) g^{-1}= \Phi(Z',X',T')$.  The space of weakly-maximal Lie subalgebras is therefore the coset space $SO(2,2)/SO(2,1)$,
which of course is a copy of AdS${}_3$.  That is, bulk points in AdS${}_3$ label the different possible embeddings of $so(2,1)$ into $so(2,2)$.  This quotient construction is special to the vacuum state of the CFT, as in general the union of all weakly-maximal subalgebras does not cover the Lie algebra generated by all total modular Hamiltonians.

As discussed in section 3, in general one expects that the bulk metric will be invariant under transformations generated by modular Hamiltonians. We saw that the total modular Hamiltonians for the vacuum of the CFT act on bulk fields as Lie derivatives everywhere in the bulk. It is thus reasonable to expect that the bulk metric will obey\footnote{This is a much stronger statement than was possible in the non-vacuum case, where the similar equation (\ref{MetricInvariant}) only held on the extremal surface.}
\begin{equation}
{\cal L}_{\xi} g_{\mu \nu} =\xi^{\alpha}\partial_{\alpha} g_{\mu \nu}+ \partial_{\mu}\xi^{\alpha}g_{\alpha \nu}+ \partial_{\nu}\xi^{\alpha}g_{\mu \alpha}=0.
\label{liederg1}
\end{equation}
This is indeed satisfied by the empty AdS metric for any of the vector fields (\ref{genxi}) associated with a vacuum modular Hamiltonian as in (\ref{xiphi}) and  (\ref{xivi}). In fact requiring the invariance (\ref{liederg1}) under the vector fields (\ref{genxi}) fixes the bulk metric to be that of empty AdS.

Alternatively one could use the generally-applicable condition  (\ref{liederg}), namely that on the extremal surface the metric should obey at each point
\begin{equation}
\Lambda_{\mu}{}^{\alpha}(\vec{X},Z,T)g_{\alpha \nu}(\vec{X},Z,T)+ \Lambda_{\nu}{}^{\alpha}(\vec{X},Z,T)g_{\mu \alpha}(\vec{X},Z,T)=0
\end{equation}
where $\Lambda_{\mu}{}^{\nu}=\partial_{\mu}\xi^{\nu}|_{\rm RT\ surface}$. The solution to this equation with $\xi^{\mu}$ given by (\ref{genxi}) is
\be
g_{\mu \nu}=\Omega(Z,X,T) \eta_{\mu \nu}.
\ee
From solving the intersecting modular Hamiltonian equations we know that the equation for the extremal surface is $Z^2=(y_2-X)(X -y_1)$ \cite{Kabat:2017mun}.
In the flat metric this surface has mean extrinsic curvature in the normal spatial direction $h_{(1)} = 1/R$ where $R$ is the radius of the circle and mean extrinsic curvature
$h_{(2)} = 0$ in the normal time direction. In the correct metric it should have zero mean curvature in all directions.  To achieve this we use (\ref{extremcon}) which fixes
the conformal factor to be that of AdS,
\[
\Omega(Z,X,T)=\frac{l^2}{Z^2}
\]
where $l$ is an undetermined constant.  The same computations can be done in the vacuum state of a higher-dimensional CFT and lead to the same result.

\section{Conclusions}

To summarize, in this paper we have proposed an algebraic approach to bulk reconstruction.  We considered the algebra generated by all modular Hamiltonians in the CFT and argued that the
bulk spacetime emerges as the parameter space of weakly-maximal subalgebras.  For the CFT vacuum this reproduces the standard quotient-space construction of empty AdS as a maximally-symmetric
spacetime.  Away from the vacuum it suggests a geometric notion of bulk modular flow, as a type of non-local symmetry associated with non-vacuum states.  But close to an extremal surface modular flow becomes a local geometric
operation -- a boost in the perpendicular directions -- and this allowed us to recover the bulk metric from the CFT.

There are many ways in which the construction presented here could fail.  In particular there's no guarantee that the requisite weakly-maximal subalgebras exist or that they can be assembled to
form a smooth $(d+1)$-dimensional manifold.  Such a breakdown is in fact expected whenever the CFT state does not have bulk dual.  Even if the algebraic construction goes through there are still things to check:
that the bulk theory is approximately local, and that quantum fluctuations are not too large.  The $1/N$ expansion is crucial  for these properties but may not be sufficient.  It would be very interesting to delineate
the necessary and sufficient conditions in more detail.

\bigskip
\goodbreak
\centerline{\bf Acknowledgements}
\noindent
DK thanks the Columbia University Center for Theoretical Physics for hospitality during this work. GL would like to thank the hospitality and support, during the course of this work, of the International Center for Theoretical Sciences (ICTS) through the program ``AdS/CFT at 20'',  the hospitality of the Mainz Institute for theoretical physics through the program ``Modern techniques for CFT and AdS'', and to thank useful discussions with participants of both programs. DK and GL would like to thank the  hospitality of CERN during the program ``Black holes, quantum information and space-time reconstruction'' where some of this work was developed.
The work of DK is supported by U.S.\ National Science Foundation grant PHY-1820734.  The work of GL is supported in part by the Israel Science
Foundation under grant 447/17.

\appendix
\section{Scalar fields\label{appendix:scalar}}
The modular Hamiltonian for a spherical region of radius $R$ centered around the origin in the vacuum of a CFT is given by
\be
\frac{1}{2\pi}H_{mod}=\frac{1}{2R}(Q_{0}-R^2 P_{0})
\ee
We use the convention \cite{Fradkin:1996is}
\bea
P_{a}\phi(x) &=& i\partial_{a}\phi(x) \nonumber\\
M_{ab}\phi(x) &=& \left( i(x_{a}\partial_{b}-x_{b}\partial_{a})+\Sigma_{ab} \right)\phi(x) \nonumber\\
D\phi(x) &=& i(\Delta+x^{a}\partial_{a})\phi(x)\nonumber\\
Q_{a}\phi(x) &=& \left( i(x^2\partial_{a}-2x_{a}x^{b}\partial_{b}-2\Delta x_{a})-2x^{b}\Sigma_{ab}\right)   \phi (x).
\eea
$\Sigma_{ab}$ are spin matrices that depend on the spin of the primary field $\phi(x)$ and $\Delta$ is the conformal dimension of $\phi(x)$.
The action of the modular Hamiltonian on a scalar operator of dimension $\Delta$ is given by
\be
\frac{1}{2i\pi}[H_{mod}, {\cal O}(t,\vec{x})]=\frac{1}{2R}((t^2+\vec{x}^2-R^2)\partial_{t}+2t\vec{x}\partial_{\vec{x}}+2t\Delta ){\cal O}(t,\vec{x})
\ee
We now compute the commutator of the total CFT modular Hamiltonian on the CFT representation of a bulk scalar operator. A bulk scalar operator is given by 
\be
\phi(Z,\vec{X},T=0)=c_{\Delta}\int dt'd\vec{y}\Theta\Big(\frac{Z^2+(\vec{x}' -\vec{x})^2-(t')^{2}}{Z}\Big)\Big(\frac{Z^2+(\vec{x}' -\vec{x})^2-(T-t')^{2}}{Z}\Big)^{\Delta-d}{\cal O}(t',\vec{x}')
\ee
where $c_{\Delta}=\frac{\Gamma(\Delta-d/2)}{2\pi^{d/2}\Gamma(\Delta-d+1)} $ and $\vec{x}'=\vec{X}+i\vec{y}$.  Commuting with the total modular Hamiltonian and integrating by parts one gets (for $\Delta>d$)
\bea
&&\frac{1}{2i\pi}[H_{mod},\phi(Z,\vec{x},T)]= c_{\Delta}\frac{\Delta-d}{R}\int dt'd\vec{y}\,\Theta(\sigma)(\sigma)^{\Delta-d-1}{\cal O}(t',\vec{x}')I_1\nonumber\\
&&\sigma = \frac{Z^2+(\vec{x}' -\vec{x})^2-(T-t')^{2}}{Z}\nonumber\\
&&I_1=\frac{1}{Z}\Big((t'-T)(Z^2+\vec{x}^2-R^2+T^2)+T(Z^2+\vec{X}^2-\vec{x'}^{2}+(T-t')^2)\Big)\nonumber
\eea
This can be seen to correspond to 
\begin{equation}
\frac{1}{2i\pi}[H_{mod},\phi(Z,\vec{x},T)]=\frac{1}{2R}(Z^2+\vec{x}^2-R^2+T^2)\partial_{T}\phi(Z,\vec{X},T)+\frac{T}{R}(Z\partial_{Z}+\vec{X}\partial_{\vec{X}})\phi(Z,\vec{X},T)
\label{hmodactscal}
\end{equation}

\subsection{$\Delta=d$ case}
In this case one can not ignore the $\delta(\sigma)$ one gets after integration by parts. So after integration by parts we have
\bea
\frac{1}{2i\pi}[H_{mod},\phi(Z,\vec{X},T)]=c_{d}\int dt'd\vec{y} \, \delta(\sigma)\frac{2}{Z}\left( (t'-T)(t'^2+\vec{x'}^2-R^2+2\vec{x'}(\vec{X}-\vec{x'}))+2T\vec{x'}(\vec{X}-\vec{x'}) \right)\nonumber
\eea
Because of the $\delta(\sigma)$ we can just add inside the brackets $(Z^2+(\vec{x}' -\vec{x})^2-(T-t')^{2})$ to obtain
\bea
&&\frac{1}{2i\pi}[H_{mod},\phi(Z,\vec{x},T)]= c_{d}\frac{1}{R}\int dt'd\vec{y}\,\delta(\sigma){\cal O}(t',\vec{x}')I_1\nonumber\\
&&I_1=\frac{1}{Z}\Big((t'-T)(Z^2+\vec{x}^2-R^2+T^2)+T(Z^2+\vec{X}^2-\vec{x'}^{2}+(T-t')^2)\Big)\nonumber
\eea
which once again gives (\ref{hmodactscal}).

\subsection{$\Delta=d-1$ case}

In this case
\[
\phi(Z,\vec{x},T=0)=\tilde{c}_{d-1}\int dt'd\vec{y}\,\delta \left(\frac{Z^2+(\vec{x}' -\vec{x})^2-(t')^{2}}{Z} \right){\cal O}(t',\vec{x}')
\]
where $\tilde{c}_{d-1}=\frac{\Gamma(d/2)}{(d-2)\pi^{d/2}}$. Computing the action of the modular Hamiltonian, after integrating by parts and using $\delta(\sigma)=-\sigma\delta ' (\sigma)$ we get
\[
\frac{1}{2i\pi}[H_{mod},\phi(Z,\vec{x},T)]=\frac{\tilde{c}_{d-1}}{ZR}\int dt'd\vec{y}\,\delta(\sigma){\cal O}(t',\vec{x}')\Big((t'-T)(Z^2+\vec{x}^2-R^2+T^2)+T(Z^2+\vec{X}^2-\vec{x'}^{2}+(T-t')^2)\Big)
\]
which again can be seen to be (\ref{hmodactscal}).

\section{Massive vector fields\label{appendix:massiveHmod}}

The action of the total modular Hamiltonian (\ref{mod1})  on a primary CFT current of dimension $\Delta$ is given by
\bea
\frac{1}{2i\pi}[ H_{mod}, j_{0}(t,\vec{x}) ] &=&\frac{1}{2R}\left(((t^2+\vec{x}^2-R^2)\partial_{t}+2t\vec{x} \cdot \partial_{\vec{x}} +2t \Delta )j_{0}+2\vec{x}\cdot \vec{j}\right)  \nonumber\\
\frac{1}{2i\pi}[ H_{mod},j_{i} (t,\vec{x})] &=& \frac{1}{2R}\left(((t^2+\vec{x}^2-R^2)\partial_{t}+2t\vec{x} \cdot \partial_{\vec{x}} +2t \Delta )j_{i}+2x_i j_{0} \right) 
\label{achmodspd}
\eea
Defining $j_{z}=\frac{1}{d-\Delta -1} \partial^{\mu}j_{\mu}$ one gets
\begin{equation}
\frac{1}{2i\pi}[ H_{mod}, j_{z}(t,\vec{x}) ]=\frac{1}{2R}\left(((t^2+\vec{x}^2-R^2)\partial_{t}+2t\vec{x} \cdot \partial_{\vec{x}} +2t(\Delta+1) )j_{z}+2j_{0} \right) 
\label{modjz}
\end{equation}
Note that the commutator of the total modular Hamiltonian looks like the one for the scalar case plus another term.
 
Bulk vector fields are represented in terms of CFT operators as \cite{Kabat:2012hp}
\begin{eqnarray}
ZV_{\mu} &= & \int K_{\Delta}\ j_{\mu} +\frac{Z}{2(\Delta-\frac{d}{2}+1)}\partial_{\mu}\int K_{\Delta+1}\  j_{z}\nonumber\\
V_{z} &=& \int K_{\Delta} \  j_{z}\nonumber\\
K_{\Delta} &=& \frac{\Gamma(\Delta-\frac{d}{2}+1)}{\pi^{d/2}\Gamma(\Delta-d+1)}\Theta\Big(\frac{Z^2+(\vec{x}' -\vec{X})^2-(t'-T)^{2}}{Z}\Big)\Big(\frac{Z^2+(\vec{x}' -\vec{X})^2-(t'-T)^{2}}{Z}\Big)^{\Delta-d}\nonumber
\end{eqnarray}
Note that $K_{\Delta}$ are the smearing functions used for primary scalars.
We start with the simple case of $[H_{mod}, V_{Z}]$. From (\ref{modjz}) we see that the commutator looks like that for a scalar operator of dimension $\Delta$ plus two terms,
one proportional to $j_z$ and one to $j_0$.
\bea
\frac{1}{2i\pi}[ H_{mod}, V_{Z}(Z,\vec{X},T)] &=&{\rm (scalar\ result)} + \frac{1}{2R} \left( 2\int K_{\Delta}j_{0}+2\int K_{\Delta} t' j_{z}\right)
\eea
where
\be
{\rm (scalar\ result)} = \xi^{\mu}_{R,0}\partial_{\mu} V_{Z}.
\ee
Noting that
\begin{equation}
\frac{Z}{2(\Delta-d/2+1)} \partial_{T} \int K_{\Delta+1} j_{z}=\int K_{\Delta}(t'-T) j_{z}
\end{equation}
we find that
\be
\frac{1}{2i\pi}[ H_{mod}, V_{Z}(Z,\vec{X},T)]=\xi^{\mu}_{R,0} \partial_{\mu} V_{Z} +\frac{Z}{R} V_{0}+ \frac{T}{R} V_{Z}.
\ee
Now $[H_{mod},V_i]$ involves two terms.  Each term has a contribution from the scalar-like transformation plus another part,
\be
\frac{1}{2\pi i}[H_{mod},\frac{1}{Z} \int K_{\Delta} j_i] = \frac{1}{Z}\xi_{R,0}^{\mu}\partial_{\mu} \int K_{\Delta} j_i + \frac{X_i}{R Z}\int K_{\Delta} j_{0}+ \frac{1}{R Z}\int K_{\Delta}(x'-X) j_{0} \\
\ee
and (with $\alpha=\frac{1}{2(\Delta-\frac{d}{2}+1)}$)
\be
\frac{1}{2\pi i}[H_{mod},\alpha \partial_{i} \int K_{\Delta+1} j_z]=\alpha \partial_{i} \left(\xi_{R,0}^{\mu}\partial_{\mu} \int K_{\Delta+1} j_z\right)+\frac{\alpha}{2R}\partial_{i}\int K_{\Delta+1} j_0
\ee
The last term the above expressions cancel each other.  Using this and the known expression for $\xi^{\mu}_{R,0}$ one finds
\be
\frac{1}{2i\pi}[ H_{mod}, V_{i}(Z,\vec{X},T)]=\xi^{\mu}_{R,0}\partial_{\mu} V_i +\frac{T}{R}V_i +\frac{X_i}{R} V_{0}
\label{HmodVi}
\ee
A similar but slightly longer computation also gives
\be
\frac{1}{2i\pi}[ H_{mod}, V_{0}(Z,\vec{X},T)]=\xi^{\mu}_{R,0}\partial_{\mu} V_0 +\frac{T}{R}V_0 +\frac{\vec{X}}{R} \vec{V}+\frac{Z}{R}V_Z
\label{HmodV0}
\ee
If the center of the sphere is at position $Y_{i}$, then in (\ref{HmodVi}) and (\ref{HmodV0}) one just shifts $X_i \rightarrow X_i-Y_i$ and $\xi^{\mu}_{R,0}  \rightarrow \xi^{\mu}_{R,Y_i}$.

\section{Reconstructing massive vectors in AdS${}_{3}$\label{appendix:massive}}

Our goal here is to use intersecting modular Hamiltonians to represent a massive bulk vector field in terms of the CFT.

We label the vector field perpendicular to the RT surface in the spatial direction as $V_{\perp}$, the vector field parallel to the RT surface as $V_{||}$, and the time component of the vector field as $V_{0}$. Then the total modular Hamiltonian acts as
\be
\frac{1}{2i\pi}[ H_{mod}, V_{0} ] = V_{\perp},  \ \ 
\frac{1}{2i\pi}[ H_{mod},V_{\perp} ] = V_{0}, \ \
[H_{mod},V_{||}] = 0
\ee
which is of course 
\be
\frac{1}{2i\pi}[ H_{mod}, V_{\perp} \pm V_{0} ] =\pm(V_{\perp}\pm V_{0}),  \ \ 
[H_{mod},V_{||}] = 0.
\ee

However what is perpendicular or parallel to a given RT surface at a given point depends on the RT surface. In AdS${}_3$ an RT surface and the modular Hamiltonian associated with it can be labeled by its two end points $(y_1,y_2)$. Thus the above equation is more correctly written as 
\be
\frac{1}{2i\pi}[ H^{(12)}_{mod}, (V_{\perp}\pm V_{0})^{(12)}] = \pm (V_{\perp}\pm V_{0})^{(12)},  \ \ 
[H^{(12)}_{mod},V^{(12)}_{||}] = 0
\ee
Imagine we have another RT surface labeled by $(y_3,y_4)$ which crosses the RT surface labeled by $(y_1,y_2)$. At the intersection point the parallel and perpendicular vectors to the two RT surfaces are at some angle $\alpha$ to each other. This angle depends only on the conformal metric so can be easily computed from the results of the intersecting modular Hamiltonians for scalar operators. So we can write
\be
V_{\perp}^{(12)}=\cos \alpha  V_{\perp}^{(34)} + \sin \alpha  V_{||}^{(34)}, \ \ V_{0}^{(12)}=V_{0}^{(34)}, \ \ 
V_{||}^{(12)}=\cos \alpha  V_{\perp}^{(34)} - \sin \alpha  V_{||}^{(34)}
\ee
From this we see that
\be
\frac{-1}{4\pi^2}[ H^{(12)}_{mod} , [ H^{(34)}_{mod} , V^{(12)}_{\perp} ] ] = \cos \alpha   V^{(12)}_{\perp}, \ \
\frac{-1}{4\pi^2}[ H^{(12)}_{mod} , [ H^{(34)}_{mod} , V^{(12)}_{0} ] ] = \cos \alpha   V^{(12)}_{0}
\ee
So we can write the following equations,
\bea
&&\frac{1}{2i\pi}[ H^{(12)}_{mod}, (V_{\perp}\pm V_{0})^{(12)}] =\pm  (V_{\perp}\pm V_{0})^{(12)}  \nonumber \\
&&\frac{-1}{4\pi^2} [ H^{(12)}_{mod} , [ H^{(34)}_{mod} , (V_{\perp}\pm V_{0})^{(12)} ] ] = \cos \alpha   (V_{\perp}\pm V_{0})^{(12)}
\label{eqtosol}
\eea
which are  decoupled equations sufficient to determine $(V_{\perp}\pm V_{0})^{(12)}$ at the intersection of the two RT surfaces. However these equations determine $(V_{\perp}\pm V_{0})^{(12)}$ only up to a coefficient which can depend on the bulk spacetime coordinates and can be chosen differently for $(V_{\perp} + V_{0})^{(12)}$ and $(V_{\perp} - V_{0})^{(12)}$. To recover the correct $V_{0}^{(12)}$ and $V_{\perp}^{(12)}$ (up to the same overall coefficient) we need another condition. We will use the fact that $V_{0}$ (but not $V_{\perp}$) satisfies
\be
-\frac{1}{4\pi^2}[ H^{(34)}_{mod} , [ H^{(34)}_{mod} , V_{0}^{(12)} ] ] =  V_{0}^{(12)} 
\ee
Thus requiring that some linear combination of the solution to (\ref{eqtosol}) corresponding to $V_{0}$ obeys this, means that we can 
get the correct $V_{0}$ and $V_{\perp}^{(12)}$, up to an overall coefficient which is the same for both.
Then to get $V^{(12)}_{||}$ we use 
\be
\frac{1}{2i\pi} [ H^{(34)}_{mod} , V_{0} ]=V_{\perp}^{(34)}=\cos \alpha  V_{\perp}^{(12)} - \sin \alpha  V_{||}^{(12)}
\ee
from which with the knowledge of $V_{\perp}^{(12)}$ we can read off $V_{||}^{(12)}$ and the corresponding overall coefficient.
 
\subsection{Practicalities}
 
It remains to see how to solve (\ref{eqtosol}).
 
Is is convenient to use as independent boundary operators the combinations ${\cal O}_{\pm}={\cal O}_{\frac{\Delta \mp1}{2},\frac{\Delta \pm 1}{2}}=j_{1} \pm j_{0}$, since their commutators with the modular Hamiltonian are diagonal.
\be
\frac{y_2-y_1}{2i\pi}[H_{mod}^{(12)},{\cal O}_{\pm}]=\left (\mp (y_2+y_1)+\Delta(\bar{\xi}-\xi) \pm (\bar{\xi}+\xi) +(\xi-y_1)(y_2-\xi)\partial_{\xi}-(\bar{\xi}-y_1)(y_2-\bar{\xi})\partial_{\bar{\xi}} \right ){\cal O}_{\pm}.
\ee
We then write an ansatz
\bea
&&(V_{0} +V_{\perp})^{(12)}=\int dp dq \ f_{+}(p,q){\cal O}_{+}(p,q) + \int dp dq\  g_{+}(p,q){\cal O}_{-}(p,q),\\
&&(V_{0} -V_{\perp})^{(12)}=\int dp dq \ f_{-}(p,q){\cal O}_{+}(p,q) + \int dp dq\  g_{-}(p,q){\cal O}_{-}(p,q)
\eea
Then (\ref{eqtosol}) becomes, upon integration by parts, differential equations for $f_{\pm}(p,q)$ and $g_{\pm}(p,q)$.
Let us define differential operators ${\cal L}_{f}^{(12)},{\cal L}_{g}^{(12)}$ by
\bea
&&\frac{1}{2i\pi}\int dp dq \ f(p,q)[H_{mod}^{(12)},{\cal O}_{+}(p,q)]=\int dpdq\ ({\cal L}_{f}^{(12)} f(p,q)) {\cal O}_{+}(p,q) \\
&&\frac{1}{2i\pi}\int dp dq \ g(p,q)[H_{mod}^{(12)},{\cal O}_{-}(p,q)]=\int dpdq\ ({\cal L}_{g}^{(12)} g(p,q)) {\cal O}_{-}(p,q)
\eea
Thus
\bea
&&{\cal L}_{f}^{(12)}=\frac{1}{y_2-y_1}(-(y_2+y_1)+(\Delta -2) (p-q)+(p+q)-(q-y_1)(y_2-q)\partial_{q}+(p-y_1)(y_2-p)\partial_{p})\nonumber \\
&&{\cal L}_{g}^{(12)}=\frac{1}{y_2-y_1}((y_2+y_1)+(\Delta -2) (p-q)-(p+q)-(q-y_1)(y_2-q)\partial_{q}+(p-y_1)(y_2-p)\partial_{p}) \nonumber
\eea
With this, equation (\ref{eqtosol}) becomes
\bea
&&{\cal L}_{f}^{(12)} f_{\pm}=\pm f_{\pm}, \ \ {\cal L}_{f}^{(12)}{\cal L}_{f}^{(34)} f_{\pm}=\cos \alpha f_{\pm} \nonumber\\
&&{\cal L}_{g}^{(12)} g_{\pm}=\pm g_{\pm}, \ \ {\cal L}_{g}^{(12)}{\cal L}_{g}^{(34)} g_{\pm}=\cos \alpha f_{\pm} \nonumber
\eea
The first equation in each line is a first-order partial differential equation while the second equation is a second-order partial differential equation. However this can be simplified since
${\cal L}_{f}^{(12)}{\cal L}_{f}^{(34)}={\cal L}_{f}^{(34)}{\cal L}_{f}^{(12)}+[{\cal L}_{f}^{(12)}, {\cal L}_{f}^{(34)}]$ and $f,g$ are eigenfunctions of ${\cal L}_{f,g}^{(12)}$.
Thus we can write 
\bea
&&{\cal L}_{f}^{(12)}{\cal L}_{f}^{(34)} f_{\pm}=\cos \alpha f_{\pm} \rightarrow  ([{\cal L}_{f}^{(12)}, {\cal L}_{f}^{(34)}] \pm {\cal L}_{f}^{(34)})f_{\pm}=\cos \alpha f_{\pm} \nonumber \\
&&{\cal L}_{g}^{(12)}{\cal L}_{g}^{(34)} g_{\pm}=\cos \alpha g_{\pm} \rightarrow  ([{\cal L}_{g}^{(12)}, {\cal L}_{g}^{(34)}] \pm {\cal L}_{g}^{(34)})g_{\pm}=\cos \alpha g_{\pm} \nonumber
\eea
which are now first-order partial differential equations. Thus we have reduced the constraints on $f_{\pm}$, $g_{\pm}$ to two linear first-order partial differential equations,
just as in the scalar case.

\subsection{Example}

As an example let us solve for $f_{+}$ for the case $y_1+y_2=0$.

We start with the equation ${\cal L}_{f}^{(12)} f_{+}= f_{+}$.
The most general solution to this equation using the methods of characteristics is
\be
f_{+}(p,q)=\tilde{f}_{+}(s) ((p-y_1)(y_2-p)(y_2-q)(q-y_1))^{\frac{\Delta-2}{2}}\left (\frac {(y_2-q)(q-y_1)}{(p-y_1)(y_2-p)} \right )^{1/2} \frac{p-y_1}{y_2-p} 
\ee
where $\tilde{f}_{+}$ is any function of the variable
\be
s=\frac{(q-y_1)(p-y_1)}{(y_2-q)(y_2-p)}
\ee
One can then insert this into the second equation $([{\cal L}_{f}^{(12)}, {\cal L}_{f}^{(34)}] + {\cal L}_{f}^{(34)})f_{+}=\cos \alpha f_{+}$ where 
\be
\cos \alpha =\frac{y_{2}^{2}-y_3 y_4}{y_2 (y_4-y_3)}.
\ee
We then get an equation for $\tilde{f}_{+}(s)$.
Doing this results after some algebra in an equation
\be
{1 \over \tilde{f}_{+}} \frac{d\tilde{f}_{+}}{ds} = \frac{(\Delta-3)}{2}\frac{s-\beta}{s(s+\beta)}
\ee
whose solution is $\tilde{f}_{+}(s)=c(y_i) \left (\frac{(s+\beta)^2}{s} \right )^{\frac{\Delta-3}{2}}$ where
$\beta=\frac{\frac{1}{2}+\frac{X_{0}}{2y_2}}{\frac{1}{2}-\frac{X_{0}}{2y_2}}$, $X_{0}=\frac{y_2^{2}+y_3 y_4}{y_3+y_4}$ is the spatial coordinate of the intersection of the RT surfaces,
and $c_{+}(y_1,y_2,y_3,y_4)$ is an overall coefficient that could depend on the boundary segments.
This gives a result for $f_{+}(p,q)$, namely
\be
f_{+}(p,q)=c_{+}(y_i)(Z^2+(p-X_0)(q-X_0))^{\Delta-3}(p+y_2)^2
\ee
where $Z^2=(X_{0}-y_1)(y_2-X_{0})$ is the bulk radial coordinate of the intersection of the RT surfaces.

\section{Gauge fields\label{appendix:gauge}}

Let's see what we get by letting the modular Hamiltonian act on the CFT representation of a bulk gauge field in the gauge $A_Z =0$.

The modular Hamiltonian for a spherical region is given in (\ref{mod1}).
The action of the modular Hamiltonian on a boundary current $(j_0,\vec{j})$ is given by (\ref{achmodspd}) with $\Delta=d-1$.
The representation of the bulk gauge field is ($\vec{x}'=\vec{x}+i\vec{y}$)
\bea
ZA_{0}(Z,\vec{x},T)=\frac{1}{{\rm vol}(S^{d-1})}\int dt'd\vec{y}\,\delta \left ( \frac{Z^2+(\vec{x}' -\vec{x})^2-(t'-T)^{2}}{2Z} \right )j_{0}(t',\vec{x}') \nonumber\\
ZA_{i}(Z,\vec{x},T)=\frac{1}{{\rm vol}(S^{d-1})}\int dt'd\vec{y}\,\delta \left ( \frac{Z^2+(\vec{x}' -\vec{x})^2-(t'-T)^{2}}{2Z} \right )j_{i}(t',\vec{x}')
\label{repgauge}
\eea
Thus 
\bea
\frac{1}{2\pi i}[H_{mod}, ZA_{i}(Z,\vec{X},T)]& = &\xi^{\mu}_{R,0} \partial_{\mu}(Z A_{i}) +\frac{X_{i}}{R}\frac{1}{{\rm vol}(S^{d-1})}\int \delta(\sigma/2) j_{0} +\nonumber\\
& & \frac{1}{R}\frac{1}{{\rm vol}(S^{d-1})}\int \delta(\sigma/2)(x'_{i}-X_{i}) j_{0}
\eea
The third term is just
\be
-Z\partial_{X_{;}}\left(\frac{1}{R}\frac{1}{{\rm vol}(S^{d-1})}\int dt'd\vec{y}\, \Theta \left( \frac{Z^2+(\vec{x}' -\vec{x})^2-(t'-T)^{2}}{2Z} \right )j_{0}(t',\vec{x}')\right)\equiv -Z\partial_{i} \lambda,
\ee
where $\lambda$ satisfies
\be
\partial_{Z} \lambda=\frac{Z}{R} A_{0}(Z,\vec{X},T).
\ee
So overall one gets
\bea
\frac{1}{2\pi i}[H_{mod}, A_{i}(Z,\vec{X},T)]& = \xi^{\mu}_{R,0} \partial_{\mu} A_{i}+\frac{T}{R} A_{i}+\frac{X_i}{R}A_{0}-\partial_{i} \lambda.
\label{modga1}
\eea

The computation of $\frac{1}{2\pi i}[H_{mod}, A_{0}]$ follows a similar track. Here one needs to use conservation of the CFT current which implies
\be
\int dt'd\vec{y}\,\delta \left ( \frac{Z^2+(\vec{x}' -\vec{x})^2-(t'-T)^{2}}{2Z} \right )((t'-T)j_{0}(t',\vec{x}')+(\vec{x}'-\vec{x}) \cdot \vec{j}(t',\vec{x}'))=0
\ee
Then one finds
\bea
\frac{1}{2\pi i}[H_{mod}, A_{0}(Z,\vec{X},T)]& = \xi^{\mu}_{R,0} \partial_{\mu} A_{0}+\frac{T}{R} A_{0}+\frac{\vec{X}}{R}\vec{A}-\partial_{i} \lambda
\label{modga2}
\eea

Thus we see that the modular Hamiltonian acting on a CFT representation of a bulk gauge field in $A_Z=0$ gauge gives exactly what we would expect.
In particular it automatically generates the compensating gauge transformation needed to restore $A_Z = 0$.

In AdS${}_{3}$ gauge fields have a simple representation \cite{Kabat:2012hp}, $A_{a}(Z,X,T)=j_{a}(X,T)$. In this case one can see (since $\partial_0 j_{1}=\partial_{1}j_{0}$) that (\ref{achmodspd}) with $\Delta=1$ is equivalent to (\ref{modga1}) and (\ref{modga2}) with $\lambda(Z,X,T)=\frac{Z^2}{2R}j_{0}(X,T)$.

\section{Gravity\label{appendix:gravity}}

The action of the total modular Hamiltonian (\ref{mod1}) on the CFT stress tensor is given by
\bea
\frac{1}{2i\pi}[ H_{mod}, T_{ij}(t,\vec{x}) ] &=&\frac{1}{2R}\left(( (t^2+\vec{x}^2-R^2)\partial_{t}+2t\vec{x} \cdot \partial_{\vec{x}} +2dt)T_{ij}+2x_{i}T_{0j}+2x_{j}T_{i0}\right )  \nonumber\\
\frac{1}{2i\pi}[ H_{mod},T_{0i} (t,\vec{x})] &=& \frac{1}{2R}\left(( (t^2+\vec{x}^2-R^2)\partial_{t}+2t\vec{x} \cdot \partial_{\vec{x}} +2dt)T_{0i}+2x^{j}T_{ji}+2x_{i}T_{00}\right ) \nonumber \\
\frac{1}{2i\pi}[ H_{mod},T_{00} (t,\vec{x})] &=& \frac{1}{2R}\left(( (t^2+\vec{x}^2-R^2)\partial_{t}+2t\vec{x} \cdot \partial_{\vec{x}} +2dt)T_{00}+4x^{j}T_{j0} \right )
\label{achmodgrav}
\eea
Using ($a,b$ range over $0,1\cdots d-1$)
\begin{equation}
Z^2h_{ab}=\frac{d\Gamma(d/2)}{2\pi^{d/2}}\int dt'd\vec{y}\,\Theta \left ( \frac{Z^2+(\vec{x}' -\vec{X})^2-(t'-T)^{2}}{2Z} \right )T_{ab}(t',\vec{x}')
\end{equation}
one finds
\bea
&& \frac{1}{2i\pi}[ H_{mod}, Z^2h_{ij}(Z,\vec{X},T) ] = \xi^{\mu}_{R,0} \partial_{\mu} (Z^2 h_{ij})+ \frac{X_i}{R}\int K_d T_{0j}+\frac{X_j}{R}\int K_d T_{0i} + \nonumber\\
&& \hspace{5cm} \frac{1}{R}\int K_d (x'_i-X_i)T_{0j}+\frac{1}{R}\int K_d (x'_j-X_j)T_{0i} \nonumber\\
&& K_{d} = \frac{d\Gamma(d/2)}{2\pi^{d/2}}\Theta \left ( \frac{Z^2+(\vec{x}' -\vec{X})^2-(t'-T)^{2}}{2Z} \right)
\eea
Let us define (with $\vec{x}'=\vec{X}+i\vec{y}$)
\be
\epsilon_{a}=\frac{d\Gamma(d/2)}{2\pi^{d/2}}\int dt'd\vec{y}\,\Theta \left ( \frac{Z^2+(\vec{x}' -\vec{X})^2-(t'-T)^{2}}{2Z} \right )(Z^2+(\vec{x}' -\vec{X})^2-(t'-T)^{2}) T_{0a}(t',x')
\ee
which satisfies
\be\partial_{Z} \epsilon_{a}=2Z^3 h_{0a}.
\ee
Then the result above can be written as
\be
\frac{1}{2i\pi}[ H_{mod}, h_{ij}(Z,\vec{X},T) ] =  \xi^{\mu}_{R,0} \partial_{\mu} h_{ij}+ \frac{2T}{R} h_{ij}+\frac{X_j}{R}h_{i0}+\frac{X_i}{R}h_{0j}-\frac{1}{2R Z^2}(\partial_{i }\epsilon_j +\partial_{j}\epsilon_i)
\label{hijmod}
\ee

Next we consider
\bea
\frac{1}{2i\pi}[ H_{mod}, Z^2h_{0i}(Z,\vec{X},T) ] &=&  \xi^{\mu}_{R,0} \partial_{\mu} (Z^2 h_{0i})+ \frac{X_i}{R}\int K_d T_{00}+\frac{X^{j}}{R}\int K_d T_{ji} + \nonumber\\
& & \frac{1}{R}\int K_d (x'_i-X_i)T_{00}+\frac{1}{R}\int K_d (x'^j-X^j)T_{ji} \nonumber
\eea
To identify the last term we use conservation of the stress tensor
\be
\int dt'd\vec{y}\,\Theta \left ( \frac{Z^2+(\vec{x}' -\vec{X})^2-(t'-T)^{2}}{2Z} \right)(Z^2+(\vec{x}' -\vec{X})^2-(t'-T)^{2}) (-\partial_{0}T_{0i}(t',x')+\partial_{j}T_{ji}(t',x'))=0
\ee
Then integration by parts gives
\be
\frac{1}{R}\int K_d (x'^{j}-X^{j})T_{ji} =-\frac{1}{2R}\partial_{T} \epsilon_{i}
\ee
so overall we have
\be
\frac{1}{2i\pi}[ H_{mod}, h_{0i}(Z,\vec{X},T) ] =  \xi^{\mu}_{R,0} \partial_{\mu} h_{0i}+ \frac{2T}{R} h_{0i}+\frac{X_i}{R}h_{00}+\frac{X^j}{R}h_{ji}-\frac{1}{2R Z^2}(\partial_{i }\epsilon_0 +\partial_{0}\epsilon_i)
\label{h0imod}
\ee
A very similar computation  gives
\be
\frac{1}{2i\pi}[ H_{mod}, h_{00}(Z,\vec{X},T) ] =  \xi^{\mu}_{R,0} \partial_{\mu} h_{00}+ \frac{2T}{R} h_{00}+\frac{2X^j}{R}h_{j0}-\frac{1}{R Z^2}\partial_{0 }\epsilon_0
\label{h00mod}
\ee

In AdS$_{3}$ a metric perturbation has a simple representation  \cite{Kabat:2012hp}, $h_{ab}(Z,X,T)=T_{ab}(X,T)$. Using the fact that the stress tensor is traceless and conserved (\ref{achmodgrav}) is equivalent to (\ref{hijmod}), (\ref{h0imod}), (\ref{h00mod}) with $\epsilon_{a}=\frac{Z^4}{2}T_{a0}$.

\providecommand{\href}[2]{#2}\begingroup\raggedright\endgroup

\end{document}